\documentclass[conference]{IEEEtran}
\usepackage[top=1.845cm, bottom=2.88cm, left=1.5875cm, right=1.5875cm]{geometry}
\IEEEoverridecommandlockouts

\usepackage{cite}
\usepackage{amsmath,amssymb,amsfonts}
\usepackage{algorithmic}
\usepackage{graphicx}
\usepackage{textcomp}
\usepackage{xcolor}
\usepackage{subfigure}
\usepackage{orcidlink}
\def\BibTeX{{\rm B\kern-.05em{\sc i\kern-.025em b}\kern-.08em
    T\kern-.1667em\lower.7ex\hbox{E}\kern-.125emX}}
\begin{document}

\title{Demo: a Hybrid Semantic RAN Protocol Stack Design for 6G System and Its Implementation}

\author{\IEEEauthorblockN{Luhan Wang$^{~\orcidlink{0000-0002-7056-5416}}$, Haiwen Niu$^{~\orcidlink{0000-0003-3115-9463}}$, \textit{Graduate Student Member, IEEE}, Zhaoming Lu$^{~\orcidlink{0000-0002-0182-1770}}$, \\
Xiangming Wen$^{~\orcidlink{0000-0003-2793-6696}}$, \textit{Senior Member, IEEE}}
\IEEEauthorblockA{
		Beijing Laboratory of Advanced Information Networks,\\
		Beijing Key Laboratory of Network System Architecture and Convergence, \\
		Beijing University of Posts and Telecommunications, Beijing, China 100876. \\
		Email: \{wluhan, haiwenniu, lzy0372, xiangmw\}@bupt.edu.cn}
	\thanks{This work was supported by the National Key Research and Development Program of China under Grant 2024YFE0200300.} 
}

\maketitle

\begin{abstract}
Recently, Semantic Communication (SC) has been recognized as a crucial new paradigm in 6G, significantly improving information transmission efficiency. However, the diverse range of service types in 6G networks, such as high-data-volume services like AR/VR/MR and low-data-volume applications requiring high accuracy, such as industrial control and data collection, presents significant challenges to fully replacing the fundamental technologies with SC. Therefore, we design a Hybrid Semantic Communication Ratio Access Network (HSC-RAN) protocol stack demo for 6G systems to achieve compatibility and smooth transition between SC and non-SC. Specifically, we take the Physical Downlink Shared Channel (PDSCH) as an example, to efficiently integrate SC with Orthogonal Frequency Division Multiplexing (OFDM). Furthermore, we introduce a novel Downlink Control Information (DCI) format that jointly supports SC and non-SC, enabling real-time video transmission via SC and text transmission through non-SC. Experimental results demonstrate that our approach allows simultaneous transmission of semantic and non-semantic information while maintaining high-quality reconstruction at the receiver.
\end{abstract}


\section{Introduction}
\IEEEPARstart{G}{uided} by the Shannon information theory \cite{shannon1948mathematical}, conventional communication systems have been designed using the hierarchical and module separation approach, which optimize each module independent of each other. In this case, this modular separation approach, while theoretically flexible, leads to suboptimal performance in practical finite-length regimes due to error propagation across layers and accumulated quantization losses, which suffers from severe challenges of future sustainable development and cannot achieve significant performance gains within tolerable costs \cite{yang2022semantic}. 

Recently, Deep Learning (DL)-enabled Semantic Communications (SCs) have attracted widespread attention due to the superior performance and robustness than the conventional bit-driven digital communication paradigm \cite{xie2021deep}. Based on the Joint Source-Channel Coder (JSCC), semantic transmitters can map information source symbols of different modalities into continuous complex-valued channel input symbols \cite{bourtsoulatze2019deep}. Through end-to-end training and cross-layer optimization, SCs can alleviate the ``cliff-effect" in an extremely poor wireless channel environment \cite{wang2022wireless}. Compared to the bit-driven digital communication paradigm, semantic communications mainly focus on how to transmit information meanings accurately. For instance, the authors in \cite{mao2024gan} have proposed a generative adversarial network (GAN)-based semantic communication framework for text transmission without CSI feedback. For wireless image transmission, the authors in \cite{niu2024sinr} have used a Transformer-convolutional neural network (CNN) mixture block to capture non-local and local semantic information to achieve superior image reconstruction performance. Moreover, in \cite{niu2023deep}, the authors have proposed a novel DL-enabled video semantic communication system, which uses a lightweight U-Net to eliminate physical channel noise and visual semantic noise in video semantic transmission. Nevertheless, existing works consider SCs independent of the current wireless network architecture and protocol stack, limiting their large-scale promotion. 

To fill this gap, we design a Hybrid Semantic Communication Ratio Network (HSC-RAN) protocol stack demo for 6G systems, which investigates the aggregation of SC and non-SC. In detail, the main contributions of this paper are summarized as follows: 
\begin{itemize}
	\item We design a HSC-RAN protocol stack demo for 6G systems, which includes a service classification layer and a semantic layer for SC, enhancing the Medium Access Control (MAC) layer and the physical layer (PHY) in 5G networks. The protocol stack allows for the simultaneous transmission of semantic and non-semantic information, facilitating a smooth transition between SC and non-SC.
	\item We introduce a fusion mechanism to integrate SC with the Physical Downlink Shared Channel (PDSCH) and Orthogonal Frequency Division Multiplexing (OFDM), allowing efficient transmission of semantic information.
	\item We introduce an extended Downlink Control Information (DCI) format with a "ResourceType" field to support semantic and non-semantic communications, facilitating resource allocation and parsing at the receiver.
\end{itemize}

\section{Demo Description}
\begin{figure}[tbp]
	\centering
	\includegraphics[width=\linewidth]{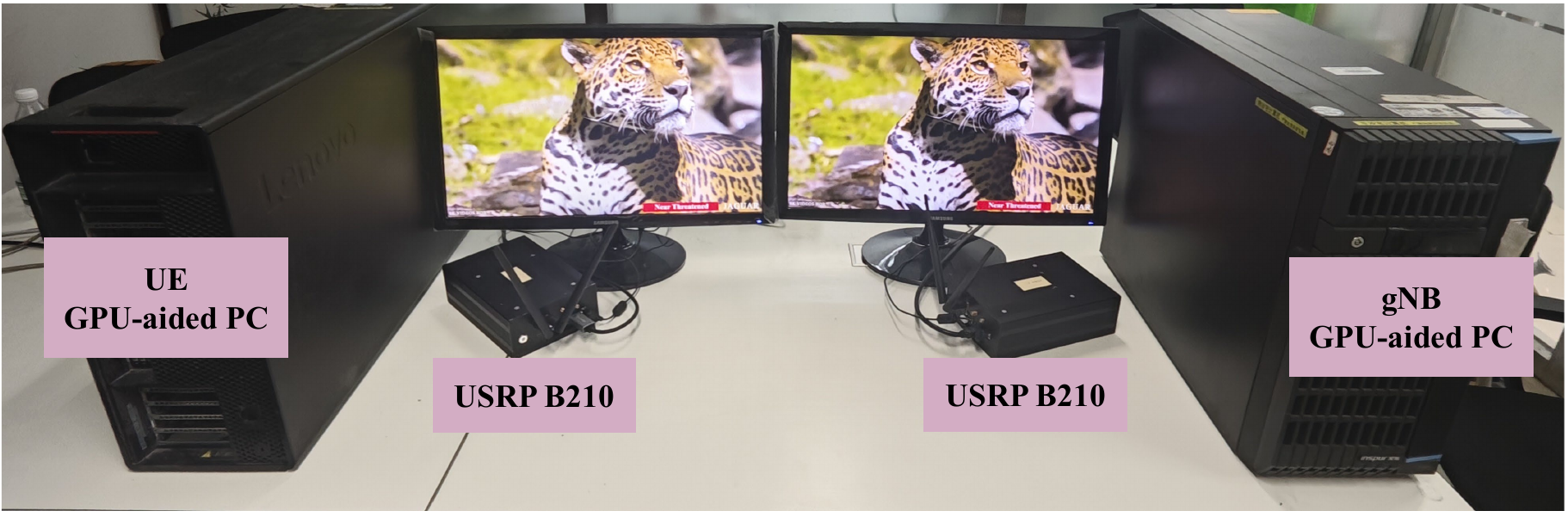}
	\caption{A hybrid semantic communication RAN protocol stack system. This demo system runs on two GPU-aided PC. }
	\label{real_system}
\end{figure}
In this demo, we utilize two Personal Computers (PC) to simulate gNB and User Equipment (UE) respectively, which can be seen in Fig. \ref{real_system}. Both PCs are equipped with Ubuntu 18.04 operating system, CUDA 11.6, PyTorch 1.13.1, and an NVIDIA A100-40GB Graphics Processing Unit (GPU). Each PC is connected to a USRP B210 software-defined radio (SDR) device.  Via these SDR devices, we can conduct practical verification of the designed scheme.

\begin{figure}[htbp]
	\centering
	\includegraphics[width=\linewidth]{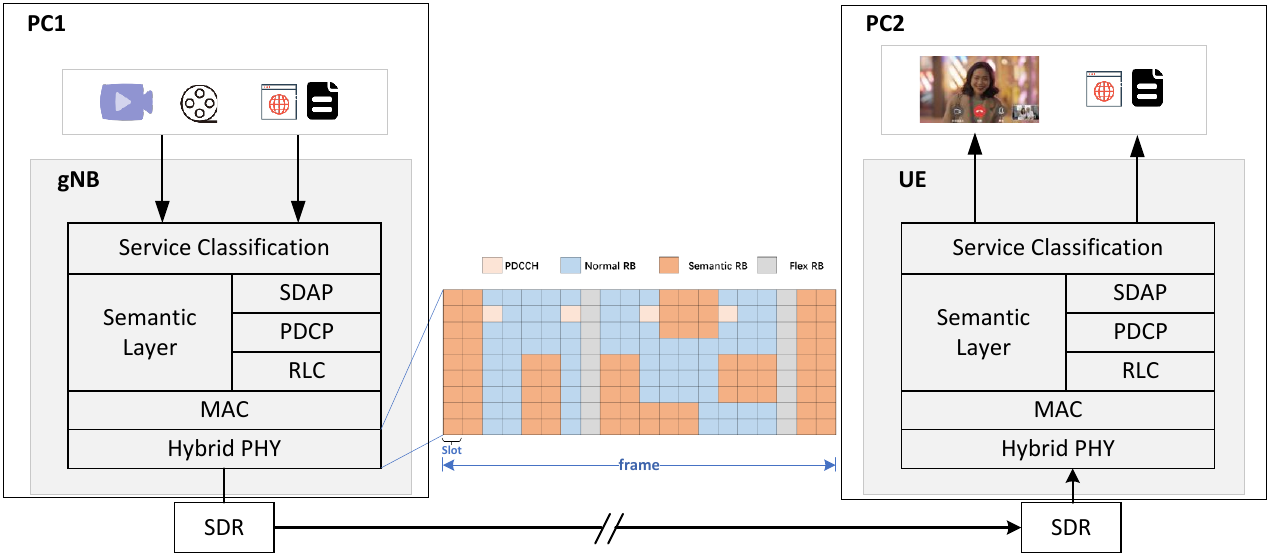}
	\caption{The overall architecture of HSC-RAN demo}
	\label{architecture}
\end{figure}
The overall architecture of the HSC-RAN demo is illustrated in Fig. \ref{architecture}.  Compared with the conventional 5G architecture, we introduce a service classification layer and a semantic layer for SC. To improve the quality of the semantic UE experience further, we enhance the original MAC layer and PHY in 5G networks. For the time being, we will keep the protocol functions above the MAC layer in traditional communication unchanged. Moreover, we divide ratio Resource Blocks (RBs) into semantic and non-semantic RBs. Sequentially, we use semantic RBs to transmit real-time video collected by the camera and non-semantic RBs to transmit text, verifying the feasibility of the HSC-RAN.
\subsection{Fusion mechanism for SC and PHY}
\begin{figure}[htbp]
	\centering
	\includegraphics[width=\linewidth]{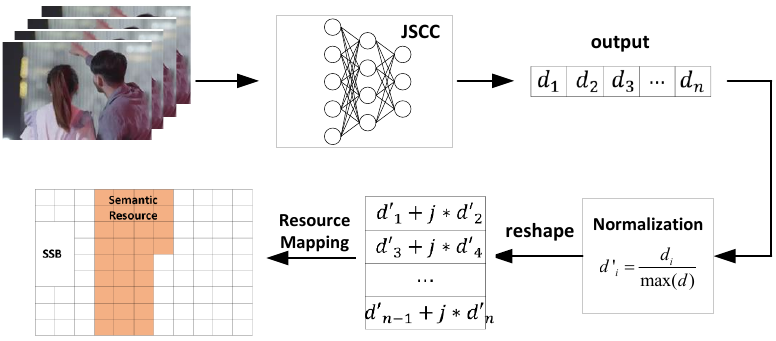}
	\caption{The fusion mechanism between SC and PHY}
	\label{fusionmech}
\end{figure}
In this demo, we design a fusion mechanism for SC and PHY, as shown in Fig. \ref{fusionmech}. Firstly, we directly input the collected video frame by frame into a DL-enabled JSCC to obtain the semantic vectors. Then, the semantic vectors are normalized and reshaped as a row vector. After power normalization, adjacent elements in the row vector are taken as the real and imaginary parts of the constellation points, respectively. Finally, we map the constellation points onto the allocated RBs. Indeed, we have retained the primary synchronization signal (PSS) used for synchronization in traditional communication. At the receiver, the reversed operations will be processed.

\subsection{Protocol workflow}
\begin{figure}[htbp]
	\centering
	\includegraphics[width=\linewidth]{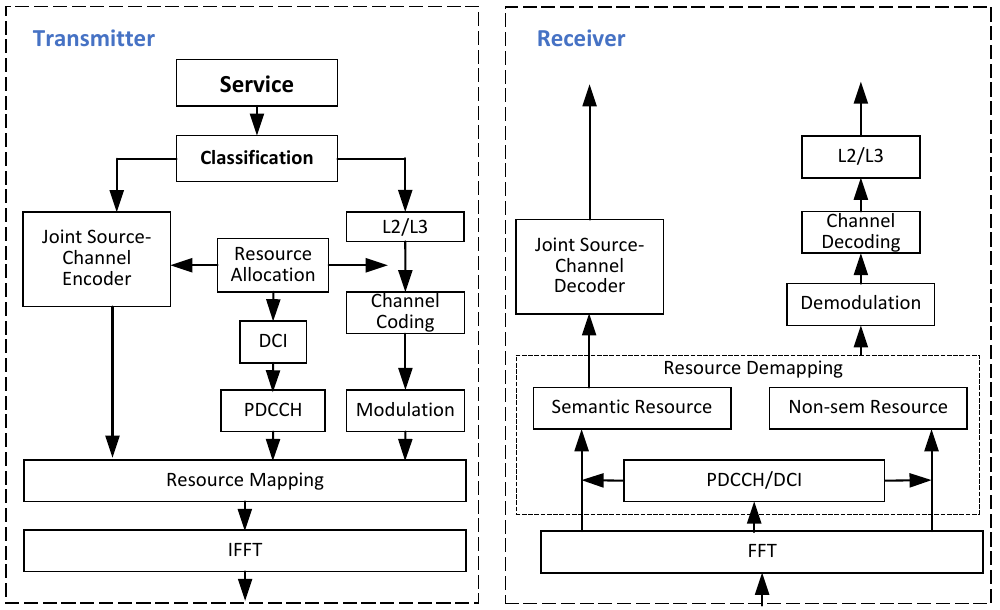}
	\caption{The overall workflow of HSC-RAN demo}
	\label{processflow}
\end{figure}
The HSC-RAN protocol stack process is shown in Fig. \ref{processflow}. The service classification layer can automatically decide whether to use SC or non-SC based on service characteristics. At the transmitter, the SC layer is responsible for encoding and decoding the corresponding service data based on the DL-enabled JSCC, while the MAC layer allocates RBs based on service demands and resource availability. In addition, we extend the DCI in the PDCCH by introducing the ``ResourceType" field in TABLE. \ref{ResourceType} to indicate semantic and non-semantic resources.  Indeed, some semantic-aware resource allocation methods \cite{cheng2023resource, ding2023joint, yan2022resource} can be deployed. Sequentially, the conventional OFDM will be executed. At the receiver, resource demapping is performed based on the DCI, and the appropriate parsing method is selected according to the service type.
\begin{table}[htbp]
	\centering
	\caption{The introduced ``ResourceType" field}
	\label{ResourceType}   
	\footnotesize
	\renewcommand{\arraystretch}{1.3}
	\setlength{\tabcolsep}{1.3mm}
	\begin{tabular}{|c|c|c|}
		\hline
		\textbf{FieldItem} & \textbf{Bits} &  \textbf{Reference}  \\
		\hline
		ResourceType & 1 & 0: Non-semantic, 1: Semantic \\
		\hline
	\end{tabular}
\end{table}

\subsection{DL-enabled JSCC for wireless video transmission}
\begin{figure}[htbp]
	\centering
	\includegraphics[width=\linewidth]{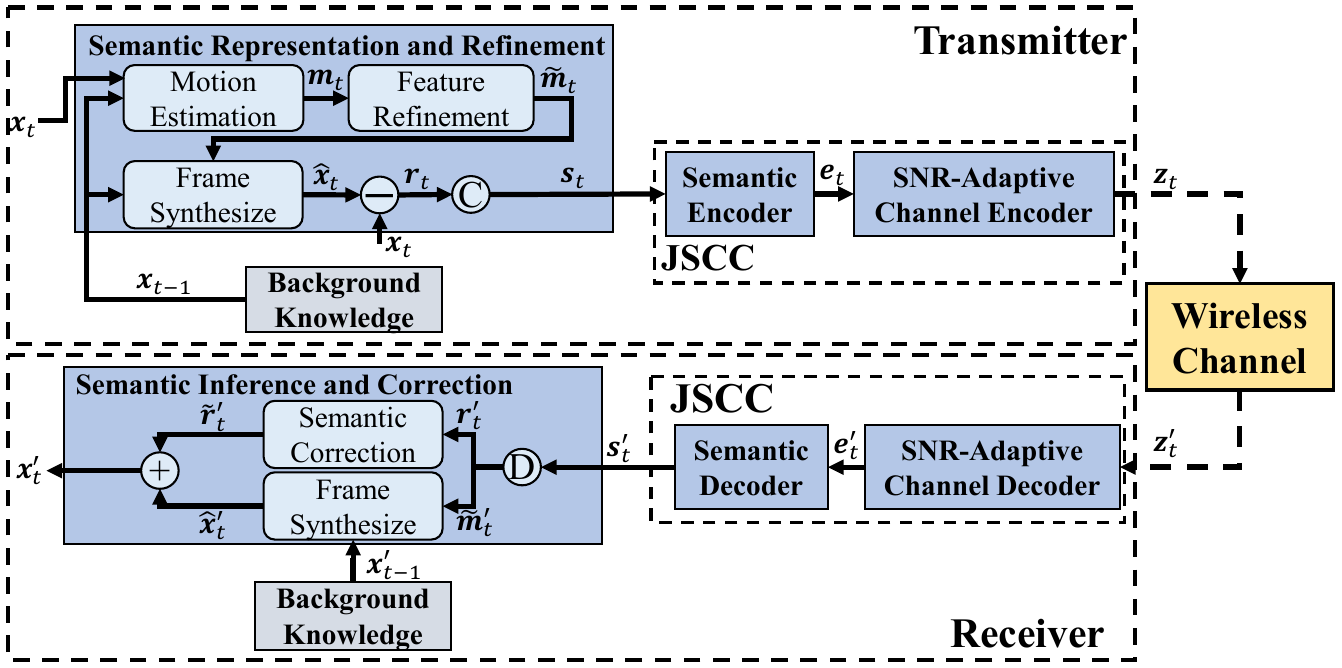}
	\caption{The architecture of DL-enabled JSCC for video transmission}
	\label{JSCC}
\end{figure}

In this section, we briefly introduce the designed JSCC for video transmission. As shown in Fig. \ref{JSCC},  the transmitter starts with a motion estimation module to predict the motion vectors (MV) between consecutive video frames. The MV is then refined through a feature refinement module based on an encoder-decoder architecture. This module can improve the accuracy of MV estimation. The refined MV is used to synthesize the current frame, which is then transmitted as part of the semantic information. To adapt to varying channel conditions, an SNR-adaptive encoder is employed, which follows the CBAM module in \cite{woo2018cbam}. This encoder uses a channel-spatial attention mechanism to allocate different contributions to semantic information based on the current SNR.  At the receiver, a corresponding decoder reconstructs the video frame by leveraging the previous frames and the transmitted information. The semantic correction module uses a lightweight U-Net structure to refine the received data, mitigating the effects of visual semantic noise. During the deployment, the input SNR can be set as $0$ without feedback.

\subsection{Experimental results and analysis}
\begin{figure}[htbp] 
	\centering
	\subfigure[Original video frame]{
		\includegraphics[width=0.22\textwidth]{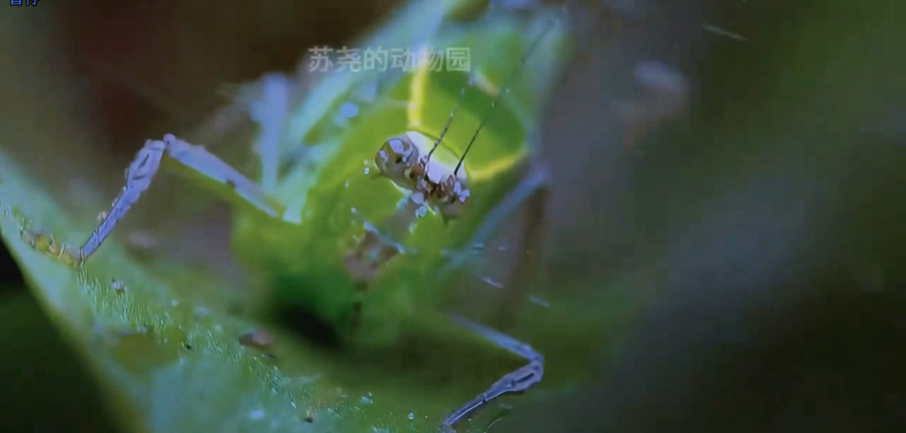}
	}
	\subfigure[Received video frame]{
		\includegraphics[width=0.22\textwidth]{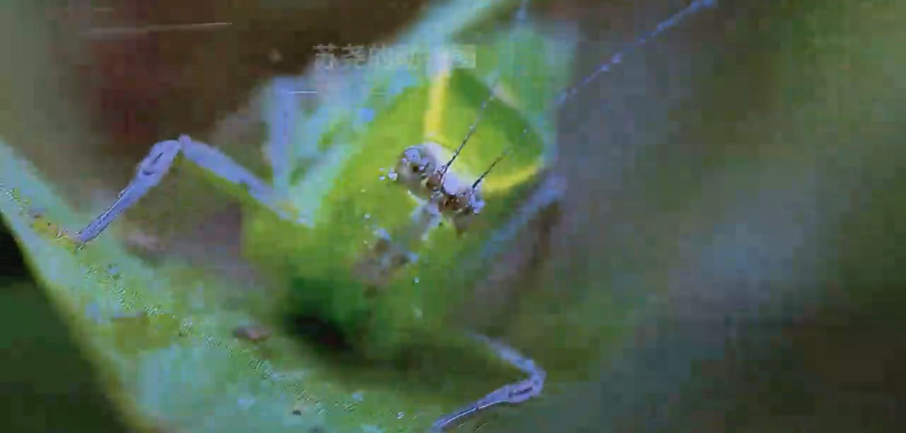}
	}
	\caption{Visual results under an actual wireless channel environment.}
	\label{visual_example}
\end{figure}

In this paper, the following experimental results are based on the SC model WITT in \cite{yang2023witt}. We regard each frame in the test video as an intra-coded picture (I-frame or keyframe). The visual results can be seen in Fig. \ref{visual_example}. The Channel Bandwidth Ratio (CBR) is $0.0417$ approximately for each I-frame, which indicates the ratio of the number of constellation points to the number of pixels. Under an actual wireless channel environment, the average test Peak Signal-to-Noise Ratio (PSNR) is $26.34$dB and the average test Multi-Scale Structural Similarity (MS-SSIM) is $0.9773$. At IEEE WCNC, we plan to deploy a camera to capture the on-site environment, and transmit the real-time video via SC and chat texts through non-SC.  

\section{Conclusion}
This paper presents a HSC-RAN protocol stack design for 6G systems, aiming to achieve compatibility between SC and traditional non-SC methods. The authors implemented a real-time video transmission system using SC in an actual wireless environment, demonstrating its feasibility. They introduced a service classification layer and a semantic layer to the protocol stack, enhancing the MAC and PHY layers to support both SC and non-SC. The system uses semantic resource blocks for transmitting real-time video and non-semantic resource blocks for text, verifying the feasibility of the HSC-RAN. The experiment showed that the system could maintain high-quality reconstruction of semantic information at the receiver, highlighting the potential of SC for future 6G networks. This work is significant for promoting the large-scale adoption of SC in next-generation communication systems.

\bibliographystyle{IEEEtran}  
\bibliography{IEEEabrv, ref}

\end{document}